\documentclass[12pt,twoside]{article}   %For printing on two sides of page
\usepackage[super,sort,comma]{natbib}
\usepackage{amsmath,amsfonts,amssymb}
\usepackage[pdftex]{graphicx}  
\usepackage{caption}
\usepackage{rotating}
\usepackage{tabularx}
\usepackage[colorlinks=true,linkcolor=blue,citecolor=blue]{hyperref}
\usepackage{array,multirow,graphicx}
\usepackage[inline,shortlabels]{enumitem} %for inline enumeration
\usepackage{xcolor,soul}

\usepackage{tikz}
\usetikzlibrary{arrows,intersections}
\usetikzlibrary{math} %needed tikz library

\usepackage{fancyhdr}		%Gives headers and footers defined below
\usepackage[section]{placeins}   %
\makeatletter \renewcommand\@biblabel[1]{$^{#1}$} \makeatother
\newcommand{\cen}[1]{\begin{center} #1 \end{center}}

\usepackage[mathlines]{lineno}
\usepackage{hyperref}
\hypersetup{ colorlinks,
    citecolor=blue,
    filecolor=blue,
    linkcolor=blue,
    urlcolor=blue
}

\usepackage{xcolor}
\definecolor{gray}{rgb}{0.6,0.6,0.6}
\definecolor{red}{rgb}{0.85,0,0}
\definecolor{green}{rgb}{0,0.85,0}
\definecolor{blue}{rgb}{0,0,0.85}
\definecolor{beige}{rgb}{0.92,0.87,0.78}
\usepackage[all]{hypcap}    %causes link to figures to go to figure, not caption
\usepackage{subfig}
\usepackage{floatrow}
\begin{document}

%\cen{\sf {\Large {\bfseries Quantification of Plan Aperture Modulation in Radiotherapy Treatment Plans}}} 
\cen{\sf {\Large {\bfseries Quantification of the aperture modulation in radiotherapy treatment plans}}} 
% A new metric to quantify the aperture modulation of radiotherapy treatment plans (but it's more than one metric, also MD...)
\vspace*{10mm}
%Author names here} \\
\begin{flushleft}

Victor Hernandez$^{1}$ \\
Department of Medical Physics, Hospital Sant Joan de Reus, IISPV, 43204 Tarragona, Spain \\
Universitat Rovira i Virgili (URV), Tarragona, Spain \\
% ORCID 0000-0003-3770-8486
\vspace*{3mm}

Iñigo Lara-Aristimuño \\
Department of Medical Physics, Hospital Sant Joan de Reus, IISPV, 43204 Tarragona, Spain \\
Universitat Rovira i Virgili (URV), Tarragona, Spain \\
% ORCID 0009-0007-3938-7694
\vspace*{3mm}

Ruben Abella \\
Department of Medical Physics, Hospital Sant Joan de Reus, IISPV, 43204 Tarragona, Spain \\
% ORCID 0000-0003-3770-8486
\vspace*{3mm}

Jordi Saez \\
Department of Radiation Oncology, Hospital Cl\'inic de Barcelona, 08036 Barcelona, Spain \\
% ORCID 0000-0003-4888-9323

\vspace*{5mm}
\end{flushleft}

\noindent $^{1}$ Author to whom correspondence should be addressed. \\
email: victor.hernandez@urv.cat \\
Department of Medical Physics, Hospital Sant Joan de Reus \\ 
Avinguda del Doctor Josep Laporte, 2, \\
Reus, 43204 Tarragona, Spain \\ 
\\
\noindent Conflict of Interest: None \\

%Version typeset \today\\

\pagenumbering{roman}
\setcounter{page}{1}
\pagestyle{plain}

\newpage     %may or may not be needed
\begin{abstract}
\noindent {\bf Background:} Many international guidelines, including those from the American Association of Physicists in Medicine (AAPM), emphasize the importance of quantifying plan modulation and recommend clearly documenting the modulation amount and range of radiotherapy treatment plans. However, there is no standardization or consensus on how to quantify plan modulation, mainly due to the limitations of existing metrics.\\

\noindent {\bf Purpose:} This study introduces two novel metrics to quantify the modulation of  treatment plans: the Plan Aperture Modulation (PAM) and the Modulation Factor (MF). The aim of these metrics is to provide a clear and intuitive assessment of the aperture modulation in radiotherapy plans, addressing limitations of previous metrics and facilitating integration into treatment planning systems (TPSs) and treatment planning workflows.\\

\noindent {\bf Methods:} PAM was defined as the average fraction of the target area located outside the beam aperture, weighted across all control points in a treatment plan. It was evaluated on Volumetric Modulated Arc Therapy plans for prostate with lymph nodes, lung stereotactic body radiation therapy, and head-and-neck cases. Plans with varying complexities were generated using the Eclipse TPS, and PAM was compared to established metrics including Plan Modulation (PM), Modulation Complexity Score (MCS), and monitor units per Gray (MU/Gy). MF was defined as the relative increase in MUs due to plan modulation, and the relationship between PAM and MF was investigated using analytical expressions.\\

\noindent {\bf Results:} PAM provided an intuitive and geometrically clear assessment of plan modulation and was validated across different delivery platforms, such as C-arm linacs and Halcyon systems. It outperformed the previous metrics, indicating zero modulation in Dynamic Conformal Arc plans, and demonstrated independence from confounding variables such as treatment technique, beam energy, delivery system, and patient anatomy, enabling a clear quantification of plan modulation. Derived equations allowed calculation of MF based on PAM, which is useful to assess and control the MU increase due to aperture modulation.\\

\noindent {\bf Conclusions:} PAM and MF are robust and intuitive metrics useful for quantifying modulation in radiotherapy plans. They address limitations of previous metrics and can be readily implemented in TPSs to control plan modulation during optimization and for reporting, thus facilitating improvements in treatment planning workflows and plan benchmarking in multi-institutional studies, clinical trials, and audits.\\

\end{abstract}
%\note{This is a sample note.}

\newpage     %may or may not be needed

The table of contents is for drafting and refereeing purposes only. Note
that all links to references, tables and figures can be clicked on and
returned to calling point using cmd[ on a Mac using Preview or some
equivalent on PCs (see View - go to on whatever reader).
\tableofcontents

\newpage

\setlength{\baselineskip}{0.7cm}      %double spacing		

\pagenumbering{arabic}
\setcounter{page}{1}
\pagestyle{fancy}

\section{Introduction}
%\vspace*{-3mm}

Intensity-modulated radiotherapy (IMRT) has become the standard treatment technique in radiotherapy, primarily for its ability to create highly conformal dose distributions and reduce radiation exposure to organs at risk \cite{cedric2002clinical}. IMRT works by optimizing the delivered fluences through beam aperture modulation, either by using static gantry positions or dynamic arc delivery in techniques like Volumetric Modulated Arc Therapy (VMAT)\cite{bortfeld2006imrt,otto2008volumetric}. However, the intrinsic complexity in IMRT introduces uncertainties in both calculated and delivered doses\cite{Chiavassa2019, Antoine2019}, potentially compromising treatment accuracy and requiring dosimetric verifications\cite{low2011dosimetry,Miften2018}. Moreover, it has been shown that highly complex plans are less robust to machine-related errors\cite{may2024delivery, terzidis2024impact}, limitations in TPS models\cite{brooks2024radiotherapy}, patient positioning uncertainties\cite{may2024setup}, and intrafraction motion\cite{may2025intrafraction}. Consequently, international guidelines stress the importance of quantifying plan modulation as part of patient-specific quality assurance (PSQA) for IMRT. In particular, the American Association of Physicists in Medicine (AAPM) recommends clearly documenting the modulation amount and range of the treatment plans evaluated\cite{geurts2022aapm}. Furthermore, the AAPM TG-218 report explicitly recommends replanning in cases of excessive modulation to prevent failures in treatment verifications\cite{Miften2018}. 

Despite these recommendations, a consensus has not been established regarding the quantification of plan modulation, understood as the extent to which a target projection is irradiated using smaller aperture segments\cite{Kamperis2020, Hernandez2020plan, kaplan2022plan}. The number of monitor units (MUs) or MU per dose ratios (MU/Gy) have also been used as indirect indicators of modulation, as highly modulated plans generally have smaller apertures and higher MU counts\cite{craft2007tradeoff}. However, MUs are affected by beam characteristics and patient anatomy. Various complexity metrics have also been proposed for this purpose\cite{Chiavassa2019, Antoine2019}. While these metrics offer useful insights, they fail to address all influencing factors, motivating the need for more comprehensive metrics. 

This study presents the Plan Aperture Modulation (PAM) metric and the Modulation Factor (MF), both designed to provide a clear and intuitive quantification of plan modulation. The advantages of these metrics over previously established modulation indices are investigated and discussed, and recommendations are provided for their integration into treatment planning workflows.
\vspace*{10mm}

\section{Materials and Methods}

\subsection{Metrics to quantify plan modulation}

Numerous complexity metrics have been proposed in the literature, many of which focus on different parameters or aspects of plan complexity, including plan modulation, beam aperture size and irregularity, multileaf collimator (MLC) leaf speeds, and modulation of gantry speed and dose rate, among others\cite{Chiavassa2019, Antoine2019}. In this study, we focused on the modulation of beam apertures and on the metrics Plan Modulation (PM)\cite{Du2014} and Modulation Complexity Score (MCS)\cite{McNiven2010}, as these are the ones most directly related to aperture modulation. Additionally, the ratio of monitor units to the prescribed fraction dose (MU/Gy) was also considered, as it is also used as a surrogate for plan modulation.

Beam Modulation (BM) was defined for step-and-shoot IMRT as the difference between each segment area and the total beam aperture area, averaged across all segments and weighted by MUs\cite{Du2014}. Plan Modulation (PM) represents the MU-weighted average of BM across all beams in a plan\cite{Du2014}. MCS was also developed for step-and-shoot IMRT, integrating information about both segment area and shape through the parameters Leaf Sequence Variability (LSV) and the Aperture Area Variability (AAV)\cite{McNiven2010}. MCS decreases with increasing complexity and ranges from 1 (single rectangular open beam) to 0, as smaller and irregular segments are added.  These metrics have been adapted to VMAT by considering the apertures at each control point\cite{Masi2013}.

\subsection{New metrics: Plan Aperture Modulation (PAM) and Modulation Factor (MF)}

\begin{figure*}[t!]
\vspace*{3mm}
      \includegraphics[width = 0.48\textwidth]{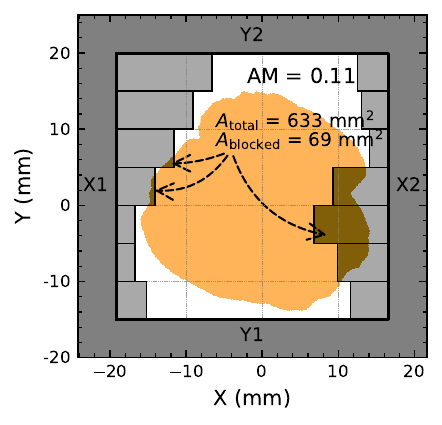}
	 \caption{Example illustrating the beam aperture and target projection in the beam’s eye view, with the total target area ($A_{\text{total}}$), the blocked area ($A_{\text{blocked}}$), and the corresponding AM value.\vspace*{3mm}}
       \label{fig:1}
\end{figure*}

The Plan Aperture Modulation (PAM) metric is designed to evaluate beam aperture modulation more directly by considering the 3D geometry of the problem. Aperture Modulation (AM) at each control point is calculated as the ratio of the blocked target projection area (outside the beam aperture) to the total target projection area:

\vspace*{-5mm}
\begin{equation}
AM = \frac{A_\text{blocked}}{A_\text{total}} \ \ ,  \label{eq:1}
\end{equation}

where $A_\text{total}$ represents the target projection area in the beam's eye view (BEV), and $A_\text{blocked}$ is the target projection area outside the beam aperture, i.e., blocked by the MLC or jaws. Figure~\ref{fig:1} illustrates an example of these concepts at a given control point.

\vspace*{3mm}

PAM is then calculated by averaging AM across all control points, weighted by the MUs delivered at each control point $j$:

\vspace*{-5mm}
\begin{equation}
PAM = \frac{\sum_{j} AM_j*MU_j }{\sum_{j} MU_j} \ \ . \label{eq:2}
\end{equation}

PAM represents the weighted average fraction of the target projection outside the beam aperture across all control points within a treatment plan. It is dimensionless and ranges from 0 (no modulation, target always fully within the beam aperture) to 1 (target fully blocked), offering a clear geometric interpretation aligned with the intuitive concept of aperture modulation. 

PAM is computed for a given treatment plan and a specific structure, leading to the possibility of computing PAM for distinct structures within the same plan. For treatment plans involving multiple dose levels, PAM can be determined separately for each target, considering the projection of each individual target structure in the BEV. The metric can also be applied at the beam level as `Beam' Aperture Modulation (BAM) by limiting the control points to those specific to a single beam.

The fraction of blocked aperture area also impacts the number of MUs of the plan. Hence, the relationship between PAM, delivered dose, and MUs was investigated following a strategy based on the estimation of the total fluence\cite{arnfield2000method, cedric1998design, saez2020novel}. The total fluence delivered to a structure by a given beam aperture at a control point $j$ is directly proportional to the number of MUs and the irradiated area receiving a given transmission:

\vspace*{-5mm}
\begin{equation}
\Psi_j=\alpha_j*MU_j*\big[A_{\text{open},j}*1+A_{\text{blocked},j}* T\big] \ , \label{eq:3} %\varphi_j
\end{equation}
\vspace*{-6mm}

where $A_{\text{open}}$ is the area of the projected structure within the beam aperture (with transmission = 1), $A_{\text{blocked}}$ is the area of the projected structure outside the beam aperture, $T$ is the average transmission of the beam collimating device (e.g., the MLC), and $\alpha$ is the proportionality constant. The units of the $\alpha$ constant are $J/m^2$ if `energy fluence' is considered, but $\alpha$ will not be used because only relative dependencies between magnitudes will be investigated. Since $A_\text{total}=A_\text{open}+A_\text{blocked}$ and using Equation~\ref{eq:1}, Equation~\ref{eq:3} can be rewritten as:

\vspace*{-8mm}
\begin{equation}
\Psi_j=\alpha_j*MU_j*\big[A_{\text{total},j}-A_{\text{blocked},j}\left(1-T \right)\big]= 
\alpha_j*MU_j*A_{\text{total},j}*\big[1-AM_j*(1-T)\big] \ \ .\label{eq:4}
\end{equation}

The average dose $D$ for a given plan can be approximated as being directly proportional to the fluence density, which can be approximated as the ratio of the total delivered fluence to the total area of the projected structure:

\vspace*{-7mm}
\begin{equation}
D_j=\beta_j\frac{\Psi_j}{A_{\text{total},j}}=k_j*MU_j*\big[1-AM_j*(1-T)\big] \ \ , \label{eq:5}
\end{equation}

where $\beta_j$ combines all the case-specific factors influencing the average dose and the constants $\alpha_j$ and $\beta_j$ were combined into $k_j=\alpha_j*\beta_j$. The constant $k_j$ has units of Gy.

The dose $D$ is then obtained by summing the contributions from all control points:

\begin{equation}
D=\sum_{j}{D_j=\sum_{j}{k_j\ast{MU}_j\ast\big[1-{AM}_j\ast(1-T)\big]}} \ \ . \label{eq:6}
\end{equation}

The distribution of $k_j$  factors depends on the patient's anatomy and plan characteristics. For a given plan, a global effective factor $k_{\text{eff}}$ can be used, resulting in:

\vspace*{-5mm}
\begin{equation}
D=k_{\text{eff}}\sum_{j}{MU_j*\big[1-AM_j*\left(1-T\right)\big]} \ \ . \label{eq:7}
\end{equation}

This $k_{\text{eff}}$ factor considers the combined effect of the various case-specific factors at each control point, described by $k_j$. In this case, rearranging terms, we can express $D$ as:

\vspace*{-5mm}
\begin{equation}
D=k_{\text{eff}}\left[MU-\sum_{j}{{{MU}_j\ast A M}_j\ast\left(1-T\right)}\right]=k_{\text{eff}}\ MU\left[1-\frac{\sum_{j}{{MU}_j\ast{AM}_j}}{MU}(1-T)\right] \ \ . \label{eq:8}
\end{equation}

From Equation~\ref{eq:2} and Equation~\ref{eq:8}, it follows that:

\vspace*{-5mm}
\begin{equation}
D=k_{\text{eff}}*MU*\big[1-PAM(1-T)\big] \ \ . \label{eq:9}
\end{equation}

Thus, Equation~\ref{eq:9} provides an approximate analytical relationship between dose, MUs, PAM, and average transmission $T$. Transmission depends on many aspects, such as the treatment unit, collimating device, and beam energy used. An average transmission value $T=0.01$ was considered in this study, which was selected based on the combined approximate transmission of the MLC with jaw tracking. Besides, expected variations around this value (e.g. $\pm 0.005$) will only have a small relative impact on Equation~\ref{eq:9}.

The Modulation Factor (MF) was defined as the relative increase in monitor units resulting from beam aperture modulation, which can be computed as the ratio of MU for a given plan to the predicted MU value for ${\rm PAM}=0$:

\begin{equation}
MF=\frac{MU}{{MU}_{PAM=0}}	 \ \ . \label{eq:10}
\end{equation}

The MF metric is dimensionless and also offers a clear interpretation.

\subsection{Cases and treatment plans evaluated}

The PAM metric was tested on various anatomical sites: prostate with lymph nodes, lung stereotactic body radiation therapy (SBRT), and head-and-neck. Prostate cases included a high-dose prostate volume (70\,Gy) and lymph nodes receiving 50.4–54\,Gy. Lung SBRT cases involved a single target prescribed 60\,Gy over 3–5 fractions. Head-and-neck cases involved three target volumes with three target volumes at 70, 59.4 and 54\,Gy over 33 fractions.

Five prostate and lung SBRT cases were selected and VMAT plans with varying complexities were generated using the Eclipse Treatment Planning System (TPS) v16.1 (Varian Medical Systems) for a Varian TrueBeam with a Millennium\,120 MLC. Prostate plans used two full arcs with 6\,MV, while lung SBRT plans employed 3–4 partial arcs with 6\,MV flattening filter-free (FFF). To generate varying complexities, plans were initially optimized with the Aperture Shape Controller (ASC)\cite{scaggion2020limiting} disabled (`Off'), then reoptimized with ASC set to `Moderate', and finally reoptimized once more by reducing MUs using the MU controller objective in the optimizer. Lung SBRT cases also included Dynamic Conformal Arc (DCA) plan, where MLCs continuously conformed to the target without intensity modulation\cite{bokrantz2020dynamic}.

Multiple dose levels in simultaneous integrated boost (SIB) techniques were evaluated using the prostate plans with lymph nodes (two dose levels) and the head-and-neck plans (three dose levels). Furthermore, to  characterize the impact of other factors, one lung SBRT plan was reoptimized using a different treatment technique (fixed-gantry IMRT) and one prostate plan was reoptimized with both a different beam energy (10\,MV) and a different delivery platform (Halcyon system with 6\,MV FFF). In all cases the same $T=0.01$ value was considered, even if the average transmission for the Halcyon system is lower\cite{saez2023universal}. Further information on the clinical cases and treatment plams used is provided as Supporting Material.

\vspace*{-5mm}
\subsection{Computations and comparisons of complexity metrics}
\vspace*{-1mm}

PM and MCS were computed using PlanAnalyzer, a MATLAB-based tool that processes DICOM RTPlan data exported from the TPS\cite{Hernandez2018}. A Python script (version 3.11) was developed to calculate the PAM metric from DICOM objects exported from the TPS, facilitating compatibility with other TPSs. The DICOM RT-structure set contours were first converted into a 3D array and meshed through the marching cubes algorithm ({\it scikit-image} library  v0.22.0). Next, the 3D mesh was projected into the BEV using the {\it open-3d} library v0.18.0, with beam aperture projections obtained from MLC and jaw positions and a resolution of 0.25\,mm. Finally, the total area ($A_\text{total}$) and blocked area ($A_\text{blocked}$) of the projected structure were obtained at each control point from the number of structure pixels inside and outside the beam aperture and the PAM metric was obtained using Equations~\ref{eq:1} and~\ref{eq:2}.

To enable the direct computation of PAM within the Eclipse TPS, a C\# program was developed using the Eclipse Scripting Application Programming Interface (ESAPI). The 3D mesh, directly available from the TPS through ESAPI, and the {\it System.Windows.Media.Media3D} library v4.0 were used to project the structure in the BEV with a resolution of 0.625\,mm. This resolution, coarser than that employed in the Python script, was selected to reduce computation times within the TPS. Finally, the beam aperture in the BEV was generated from the MLC and jaw positions, using the same resolution, and AM and PAM values were calculated. The developed codes are available upon request. % to the authors.

\vspace*{-3mm}
\section{Results}
\vspace*{-2mm}

\subsection{Plan Modulation}
%\vspace*{-3mm}

\begin{figure*}[b!]
\vspace*{3mm}
      \includegraphics[width = 0.48\textwidth]{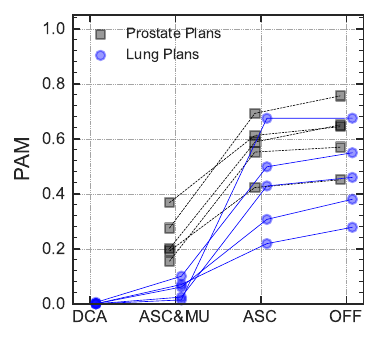}
	  \caption{PAM values for prostate (high-dose prostate target) and lung SBRT plans. Symbols represent plans with varying complexity levels, and lines connect plans for the same clinical cases. X-axis labels represent complexity control settings: OFF (no control), ASC (Aperture Shape Controller set to moderate), ASC\&MU (ASC set to moderate with MU restriction), and DCA (Dynamic Conformal Arc technique).
\vspace*{10mm}}
      \label{fig:2}
\end{figure*}

PAM results are shown in Figure~\ref{fig:2}, using the high-dose prostate volume as target volume. PAM increased with plan complexity, with the lowest values occurring when both ASC and MU restrictions were applied, indicating the least complexity. For lung SBRT cases, PAM dropped to zero for DCA plans, which lack modulation.

%\vspace*{5mm}

\begin{figure*}[t!]
%\vspace*{2mm}
      \includegraphics[width = 0.48\textwidth]{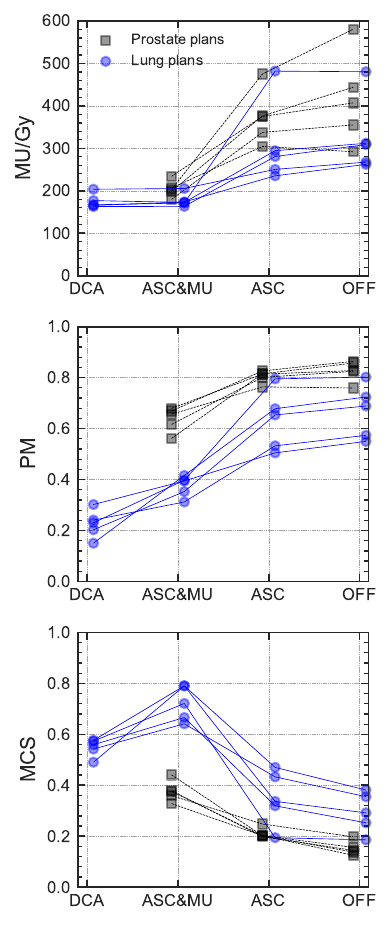}
	  \caption{Plan complexity values for prostate and lung SBRT plans for the ratio of Monitor Units to prescribed dose (MU/Gy, top row), Plan Modulation (PM, middle row), and Modulation Complexity Score (MCS, bottom row). Symbols represent plans with varying complexity levels, and lines connect plans from the same clinical cases. X-axis labels represent complexity control settings: OFF (no control), ASC (Aperture Shape Controller set to moderate), ASC\&MU (ASC set to moderate with MU restriction), and DCA (Dynamic Conformal Arc technique).  \vspace*{5mm}}
      \label{fig:3}
\end{figure*}

Figure~\ref{fig:3} shows results from other metrics. MU/Gy trends were similar to PAM but did not converge for DCA plans. PM increased with plan complexity but remained above zero for DCA plans, while MCS generally decreased with complexity. However, MCS values for DCA plans were unexpectedly lower than those for ASC\&MU, indicating higher complexity despite lacking aperture modulation.

%\vspace*{5mm}

\begin{figure*}[t!]
\vspace*{2mm}
      \includegraphics[width = 0.48\textwidth]{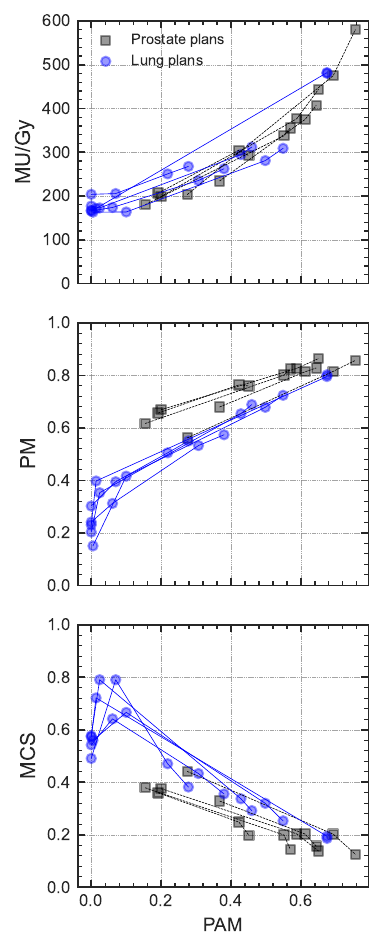}
	  \caption{Relationship between PAM values and number of Monitor Units per Gray (MU/Gy, top row), Plan Modulation (PM, middle row), and Modulation Complexity Score (MCS, bottom row). Symbols represent plans with varying complexity for lung SBRT and prostate plans, with lines connecting plans from the same clinical cases. \vspace*{5mm}}
      \label{fig:4}
\end{figure*}

Figure~\ref{fig:4} directly compares PAM with the other metrics, revealing consistent trends between metrics. The relationship between MU/Gy and PAM was particularly smooth but differences of approximately 30\% in MUs were observed for the same PAM values due to anatomical variations. PM and MCS also showed differences with respect to PAM, including abrupt changes between low PAM values (associated with VMAT plans of low complexity) and PAM = 0 values (corresponding to DCA plans).

%\vspace*{5mm}

%\clearpage

\subsection{Relationship between MU and PAM: Modulation Factor} %MU, PAM and MF

The relationship between MU/Gy and PAM was further analyzed using Equation~\ref{eq:9}. For two treatment plans (labeled `1' and `2') from the same clinical case and using the same beam setups, we hypothesized that $k_{\text{eff}}$ remains constant for both plans and cancels out. However, this approximation will likely hold only in cases where plan characteristics, such as the angular distribution of the control points weights, are similar, and requires validation. In this case, the equation simplifies to:

\vspace*{-5mm}
\begin{equation}
MU_1 * \big[1-{PAM}_1\left(1-T\right)\big]=MU_2 * \big[1-{PAM}_2(1-T)\big] \ \ . \label{eq:11}
\end{equation}

This allows the MU-PAM relationship for one plan to be derived if the values for another plan of the same case and beam setup are known. The relationship between PAM and MU can then be expressed as:

\vspace*{-5mm}
\begin{equation}
PAM_2=\frac{MU_2-MU_1\big[1-{PAM}_1\left(1-T\right)\big]}{MU_2(1-T)}	 \ \ . \label{eq:12}
\end{equation}

The reverse relationship, where MU is derived from PAM, is given by:

\vspace*{-5mm}
\begin{equation}
MU_2=MU_1 * \frac{1-PAM_1\left(1-T\right)}{1-PAM_2(1-T)}		 \ \ . \label{eq:13}
\end{equation}

\clearpage

\begin{figure*}[t!]
\vspace*{2mm}
      \includegraphics[width = 0.48\textwidth]{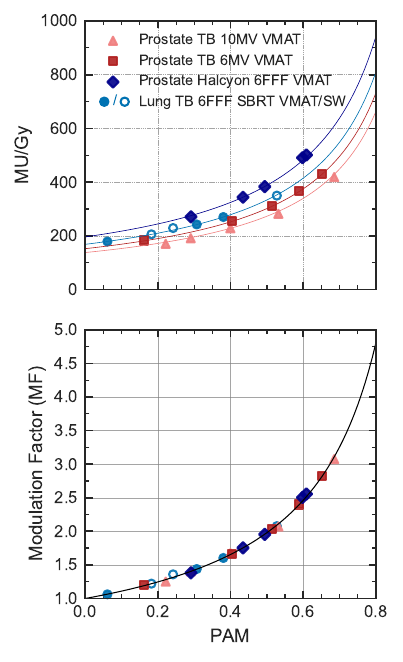}
	  \caption{Top row: Relationship between Monitor Units (MU) and PAM for a lung SBRT case with plans using different treatment techniques --Volumetric Arc Therapy (VMAT) and Sliding Windows (SW)-- and a prostate with lymph nodes case with plans using different beam energies --6\,MV, 6FFF, and 10\,MV--. Symbols represent the values for the evaluated plans and lines show the MU-PAM relationship from Equation~\ref{eq:12}. Bottom row: Modulation Factor for the same plans, with lines showing the theoretical predictions from Equation~\ref{eq:14} and symbols indicating the ratio of MU to the predicted MU for PAM\,=\,0 for each case.
\vspace*{6mm}
}
      \label{fig:5}
\end{figure*}

Figure~\ref{fig:5} (top) shows the PAM and MU/Gy values for plans from two different anatomical sites. First, a prostate case was planned using different delivery platforms (TrueBeam and Halcyon) and beam energies (6\,MV, 6\,FFF, 10\,MV). Second, plans using 6\,FFF beams were obtained for a lung SBRT case using different treatment techniques (VMAT, sliding windows). For each case, plans of various complexities were generated and compared in terms of MU/Gy and PAM values. Predictions from Equation~\ref{eq:12} were also illustrated, where plans with intermediate PAM values (the ones closest to PAM\,=\,0.4) were considered as reference (i.e., plan "1" in Equation~\ref{eq:12}). Discrepancies between computed (symbols) and predicted values (dashed lines) were minimal, validating the accuracy of the derived equations, as well as the hypothesis that a constant $k_{\text{eff}}$ can generally be assumed for plans within the same clinical case and a similar beam setup. This held true even when VMAT and sliding windows plans were used, which is why all lung SBRT plans were plotted together in Figure~\ref{fig:5} (top).

However, notable differences were obtained between plans from different treatment sites, beam energies, and delivery systems, which was expected due to the known dependencies of the number of MU. To account for case-specific variations in MUs arising from differences in treatment sites, beam energies, and delivery systems, the Modulation Factor MF was defined as the ratio of MU (or MU/Gy) for a given plan to its prediction for PAM\,=\,0 (see Equation~\ref{eq:10}). Using Equation~\ref{eq:13} and considering $\rm{PAM}_{2}=0$, the MU for PAM\,=\,0 can be obtained, and MF can be expressed in terms of PAM as:

\begin{equation}
MF=\frac{MU}{{MU}_{PAM=0}}=\frac{1}{1-PAM\ast\left(1-T\right)}	 \ \ . \label{eq:14}
\end{equation}

The bottom row in Figure~\ref{fig:5} illustrates the relationship between MF and PAM for the prostate and lung SBRT plans plotted in at top row of Figure~\ref{fig:5}, highlighting how case-specific differences were removed, making results independent of treatment site, beam energy, and delivery system. Calculated MF values (symbols) aligned closely with Equation~\ref{eq:14} predictions (dashed line), validating the proposed equations and the derived relationship between MF and PAM.

\vspace*{18mm}

\subsection{Plans with simultaneous integrated boost (SIB)}
\vspace*{-3mm}

\begin{figure*}[b!]
      \includegraphics[width = 0.95\textwidth]{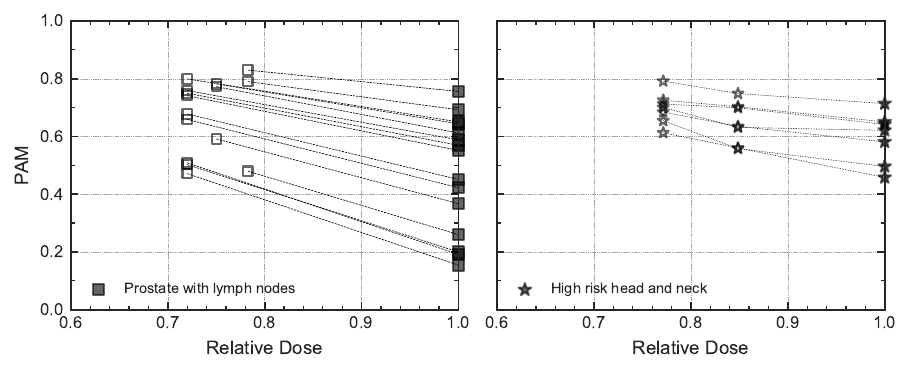}
	  \caption{PAM values for prostate lymph nodes (left) and head-and-neck plans (right) as a function of relative dose (compared to the highest-dose target). Lines connect data from the same plans, with symbols representing different target volumes.\vspace*{15mm}}
      \label{fig:6}
\end{figure*}

For IMRT plans with simultaneous integrated boost, where multiple targets receive different doses\cite{popple2005simultaneous, jolly2011rapidarc}, PAM can be computed for all different targets and will vary by structure. Figure~\ref{fig:6} shows the PAM values for the prostate and head-and-neck plans computed for all the target structures as a function of their relative dose. As can be seen, higher-dose structures exhibit lower PAM values, while lower-dose structures show higher PAM values due to a greater proportion of blocked apertures needed for reduced doses.

%\newpage

The relationship between PAM and the evaluated structure was also investigated using Equation~\ref{eq:9}, in which the number of MU cancels out because both sides of the equation correspond to the same plan, resulting in:

\begin{equation}
\frac{1-{PAM}_2(1-T)}{D_2/k_{{\rm eff}_2}}=\frac{1-{PAM}_1(1-T)}{D_1/k_{{\rm eff}_1}}  \ \ . \label{eq:15}
\end{equation}

Thus, PAM value for structure `2' can be estimated from the PAM value of structure `1', the ratio of prescribed doses, and a correction factor $C$ that depends on the specific geometry of each structure ($C=k_{\rm eff_1}/k_{\rm eff_2}$):

\begin{equation}
PAM_2=\frac{1-C*D_2/D_1*\big[1-PAM_1*(1-T)\big]}{1-T} \label{eq:16}
\end{equation}

This equation shows that PAM for a given structure varies linearly with the dose ($D_{2}$ in Equation~\ref{eq:15}), with a proportionality coefficient strongly dependent on the target-specific geometry, as reflected by the constant $C$. The head-and-neck plans shown in Figure~\ref{fig:6}, featuring three dose levels, further demonstrate this relationship, with PAM values consistently increasing as relative doses decrease. The lowest PAM values, hence, correspond to the targets receiving the highest prescribed doses.
\vspace*{3mm}

\section{Discussion}
\vspace*{-4mm}

Plan complexity metrics were initially derived from fluence maps, measuring fluence variations between adjacent pixels\cite{webb2003use, giorgia2007acceptably}. However, fluence-based metrics present inherent limitations, as they do not account for variations in segmentation that can yield identical total fluence maps\cite{Du2014}. Moreover, modern optimization algorithms focus on directly optimizing machine-specific parameters rather than fluence\cite{unkelbach2015optimization}. Consequently, fluence-based metrics have gradually been replaced by metrics that incorporate specific plan parameters, such as MLC positions\cite{Chiavassa2019, Antoine2019}. However, these newer metrics also have limitations. PM and MCS, for example, were developed for static-gantry IMRT, using the union of segments as a proxy for the target's total projection at each gantry angle\cite{McNiven2010,Du2014}. In VMAT plans, however, continuous changes in the target projection with gantry rotation complicate these metrics' applicability\cite{masi2021robotic}. 

PAM quantifies plan modulation through a geometry-based methodology, making it inherently independent of parameters such as beam energy, dose output calibration, and depth, in contrast to MUs. Moreover, PAM explicitly accounts for the BEV projection of the target and quantifies the fraction of its occluded area, thereby avoiding geometric assumptions about the total target area, as required by metrics such as MCS and PM. As a result, PAM addresses key limitations of existing metrics and provides a more consistent and reliable assessment of plan modulation.

In this study, we evaluated plans across different treatment sites (lung SBRT, prostate with lymph nodes, and head-and-neck plans with various SIB dose levels), treatment techniques (VMAT, DCA, sliding windows), planning strategies (varying levels of ASC and MUs), and delivery platforms (TrueBeam and Halcyon). The aim was to demonstrate that PAM and MF perform consistently and intuitively across these diverse clinical scenarios.

For DCA plans, for example, MCS yielded unexpectedly low values, indicating higher modulation compared to certain VMAT plans from the same clinical cases. This is likely due to limitations in MCS, which constrain its range of validity. Specifically, MCS evaluates the variations in segment areas relative to the maximum aperture defined by the union of all segments rather than the target projection in the BEV. Furthermore, MCS also quantifies relative variations in leaf positions with respect to the overall distribution of leaf positions, which differs substantially between DCA and VMAT plans.

The derived equations were used to explore the dependencies among PAM, MUs, and MF. Although these analytical expressions are based on simple assumptions, the results presented in Figure~\ref{fig:5} demonstrate good agreement between predicted and computed MU–PAM values (top row), as well as consistency between predicted and computed MF values (bottom row), thereby supporting the validity of the approach.

Figure~\ref{fig:5} (bottom row) demonstrates that MF effectively eliminates case-to-case variations in MUs in all cases, allowing for a clear interpretation. Note that MF increases with PAM with a varying slope, with steeper variations for large PAM values. For example, an increase in PAM from 0.2 to 0.3 raises the MF value by 0.14 (i.e., the number of MUs by 14\%), while increasing PAM from 0.6 to 0.7 raises MF by 0.32 (a 32\% increase in MUs). Thus, the MF metric further emphasizes the impact of high PAM values and highly-modulated plans.

The term modulation factor has been used by some authors to refer to the MU/Gy ratio \cite{burton2020robust, peiris2025effect}. In contrast, we introduced a new definition of the modulation factor (MF) to quantify the increase in MUs resulting from beam aperture modulation. This definition provides clear advantages over the MU/Gy ratio, as it eliminates influencing factors related to the beam energy, depth, and dose output calibration. By definition, the proposed MF depends on MUs and, consequently, on dose calculations. However, we demonstrated that MF can be directly determined from PAM using Equation~\ref{eq:14}, which highlights a particularly useful application of the PAM metric.

MF can also be calculated for individual beams by substituting PAM with BAM. A disadvantage of MF is that it depends slightly on the delivery system through the average transmission $T$, in contrast to PAM, but the impact of differences in $T$ values is minimal, particularly for clinical plans with PAM $<0.75$. In this study, $T=0.01$ was used, which is lower than the average MLC transmission of the Millennium\,120 MLC\cite{saez2023universal}, to account for jaw tracking, which further lowers the effective average transmission. For the Halcyon, we also used $T$\,=\,0.01, which is higher than the average transmission for this system\cite{saez2023universal}, to demonstrate that, for current treatment units with low transmission values, a common value of $T$\,=\,0.01 provides good agreement in general (as shown in Figure~\ref{fig:5}). 

In summary, PAM offers several key advantages: \vspace*{-4mm}

\begin{itemize}\setlength{\itemsep}{0pt}
\item It provides a clear geometric interpretation aligned with the intuitive concept of aperture modulation.
\item By definition, it depends solely on beam apertures and target projections, rendering it inherently applicable to any delivery system and treatment technique, irrespective of the specific collimation device used.
\item It is independent of factors such as beam energy, depth, and dose output calibration.
\item It facilitates the computation of MF.
\end{itemize}

The primary applications of PAM and MF are expected to be in controlling plan modulation during optimization processes and in reporting. Most TPSs currently control plan modulation through the number of MUs, but PAM and MF can replace MUs in this role, offering reduced sensitivity to influencing factors and facilitating the use of limits that are independent of beam energy, depth, and dose output calibration. PAM and MF also offer clear advantages for reporting the degree of plan modulation, especially in multi-institutional studies, due to their applicability across different delivery platforms and treatment techniques.

We advocate for the integration of PAM and MF metrics into TPS platforms. Until then, Equations~\ref{eq:12},~\ref{eq:13},~\ref{eq:14} can facilitate their use in treatment planning workflows. For instance, once PAM is calculated, the necessary MU adjustment to achieve a lower PAM value can be computed using  Equation~\ref{eq:13} and set as a constraint in the TPS optimizer. A script has also been developed to calculate PAM directly within the TPS, removing the need to export plans or structures.

Modern radiotherapy techniques, such as those involving multiple dose levels and simultaneous integrated boosts, present additional challenges in quantifying plan modulation. In these scenarios, PAM values tend to be lower for high-dose targets and higher for low-dose targets, which generally involve a greater proportion of blocked apertures (see Figure~\ref{fig:6}). As a result, applying an upper PAM limit to the low-dose target may inadvertently yield an excessively low PAM value for the high-dose target, which may require a certain level of modulation to achieve a clinically acceptable dose distribution. Therefore, in such cases, it is generally more effective to control plan modulation by focusing on high-dose targets, allowing for increased modulation in structures with lower prescribed doses.

Another key application are multi-institutional comparisons. Although reporting plan modulation and complexity is strongly recommended, these practices are often not followed, likely due to a lack of suitable metrics that are valid across different treatment units and institutions\cite{hansen2020radiotherapy, Hernandez2020plan, Kamperis2020,kaplan2022plan}. PAM provides a promising solution for comparing and benchmarking treatment plans across institutions, especially in clinical trials and audits.

While PAM and MF offer clear advantages, its computation requires BEV projections, which increases computational demands. The Python and C\# codes were validated using basic geometries, both yielding accurate PAM values with discrepancies of less than 0.01. Comparisons of the Python and C\# codes in treatment plans also showed good agreement (differences in PAM $<0.01$) and reasonable computation times: 4–30 seconds for Python, depending on the target size and the number of control points, and 1–5 seconds for the TPS script, which benefits from lower resolution and access to the 3D mesh via ESAPI. Modern TPS platforms often include tools for BEV projections, which can expedite computations and facilitate efficient implementation. Additionally, TPS optimizers can estimate MU during optimization, enabling PAM-based modulation control without recalculating at each iteration.

In the present study, correlations between plan complexity and PSQA outcomes were not investigated. Instead, this work builds upon existing knowledge about highly complex plans being less robust and potentially less accurate \cite{Chiavassa2019, Antoine2019, brooks2024radiotherapy, may2024delivery, may2024setup, terzidis2024impact, may2025intrafraction}. We focused on addressing the limitations of current modulation metrics and presented two novel indices, PAM and MF, which are based on clear geometric and physical principles, which facilitates their interpretation and expands their range of validity. While the study evaluated a limited number of anatomical sites and clinical cases, they served to validate the assumptions used and the derived relationships across clinical plans of various characteristics.

We introduced the PAM metric, demonstrated its advantages, and discussed its potential applications in radiotherapy, especially those related to plan optimization and reporting. However, a comprehensive exploration of all the potential applications of the proposed metrics was beyond the scope of this study and will be the focus of future work. Similarly, we focused on plan modulation and did not examine other aspects of plan complexity, such as aperture irregularity \cite{Du2014}, leaf speed and travel distance \cite{Masi2013}, or variations in dose rate and gantry speed \cite{Park2014}. Although beam aperture modulation is considered the primary component of plan complexity \cite{craft2007tradeoff, Hernandez2020plan}, these other characteristics are also relevant, and to fully characterize IMRT/VMAT plan complexity, it is necessary to complement aperture modulation quantification with metrics that capture additional plan attributes.

\section{Conclusions}
\vspace*{-2mm}

The PAM metric quantitatively assesses aperture modulation in radiotherapy treatment plans, offering a robust and intuitive geometry-based measure that addresses limitations of previous metrics. The Modulation Factor MF was defined as the relative increase in MUs attributable to the modulation of beam apertures and we demonstrated that MF can be analytically computed from PAM. The proposed PAM and MF metrics can be readily integrated into TPSs and serve as replacements for MUs in controlling plan modulation during optimization processes and for reporting. In addition, they provide a valuable tool for improving treatment planning workflows and for facilitating the comparison and benchmarking of plans in multi-institutional studies, clinical trials, and audits.
\\

\section*{Acknowledgements}
\vspace*{-3mm}
This research was funded by the Spanish Society of Medical Physics (SEFM) through research Grant PI-SEFM-2022. 

\section*{References}
%\addcontentsline{toc}{section}{\numberline{}References}
\vspace*{-10mm}

\bibliographystyle{./medphy}    %if this is installed on your system,
				    %it is not essential to have the    ./

% Note that you need to typeset once, then run bibtex, then typeset another
% two times to get the references working properly.
%\section*{Supporting information}
%Additional information is available online in the Supporting Information section.

%\newlength{\imagewidth}
%\newlength{\imageheight}
%\settowidth{\imagewidth}{\includegraphics{./figures/FigureS1_v1}}
%\settoheight{\imageheight}{\includegraphics{./figures/FigureS1_v1}}
%\begin{figure*}[t!]
%      \includegraphics[width = 1.5\imagewidth]{FigureS1_v1.pdf}
%       \caption{Experimental and calculated $\Delta \phi_\text{TG}(s)$ values for the Agility. The solid curved line represents average measured data. Dashed and dotted lines show the calculated curves from each TPS. \vspace*{3mm}}
%       \label{fig:FigS1}
%\end{figure*}
%\let\imagewidth\relax
%\let\imageheight\relax
%
%\newlength{\imagewidth}
%\newlength{\imageheight}
%\settowidth{\imagewidth}{\includegraphics{./figures/FigureS2_all_v0}}
%\settoheight{\imageheight}{\includegraphics{./figures/FigureS2_all_v0}}
%\begin{figure*}[t!]
%      \includegraphics[width = 1.5\imagewidth]{FigureS2_all_v0.pdf}
%       \caption{Experimental and calculated $\Delta \phi_\text{TG}(s)$ values for the Agility. The solid curved line represents average measured data. Dashed and dotted lines show the calculated curves from each TPS. \vspace*{3mm}}
%       \label{fig:FigS2}
%\end{figure*}
\let\imagewidth\relax
\let\imageheight\relax

\end{document}